# Learning to Generate Pseudo Personal Mobility

Peiran Li[1], Haoran Zhang[2*], Wenjing Li[3], Dou Huang[3], Jinyu Chen[1], Junxiang Zhang[4], Xuan Song[4], Pengjun Zhao[2], Shibasaki Ryosuke[1]

**Abstract**

The importance of personal mobility data is widely recognized in various fields. However, the utilization of real personal mobility data raises privacy concerns. Therefore, it is crucial to generate pseudo personal mobility data that accurately reflects real-world mobility patterns while safeguarding user privacy. Nevertheless, existing methods for generating pseudo mobility data, such as mechanism-based and deep-learning-based approaches, have limitations in capturing sufficient individual heterogeneity. To address these gaps, taking pseudo-person(avatar) as ground-zero, a novel individual-based human mobility generator called GeoAvatar has been proposed – which considering individual heterogeneity in spatial and temporal decision-making, incorporates demographic characteristics, and provides interpretability. Our method utilizes a deep generative model to simulate heterogeneous individual life patterns, a reliable labeler for inferring individual demographic characteristics, and a Bayesian approach for generating spatial choices. Through our method, we have achieved the generation of heterogeneous individual human mobility data without accessing individual-level personal information, with good quality - we evaluated the proposed method based on physical features, activity patterns, and spatial-temporal characteristics, demonstrating its good performance, compared to mechanism-based modeling and black-box deep learning approaches. Furthermore, this method maintains extensibility for broader applications, making it a promising paradigm for generating human mobility data.

**Introduction**

Personal mobility data has found wide applications in urban studies (Ahas & Mark, 2005; Wu et al., 2022), including epidemiological prediction and control (Ceder, 2021; Grantz et al., 2020; Leung et al., 2021; Manica et al., 2021), transportation and travel planning (Joubert & de Waal, 2020; Torre-Bastida et al., 2018), event simulation or disaster management (Deng et al., 2021; X. Song et al., 2015; M. Yu et al., 2018), and emission reduction (Yi & Yan, 2020; Zhang et al., 2019, 2020), among others. As a result, acquiring comprehensive human mobility data has become crucial (Shin et al., 2020). The rapid proliferation of portable positioning devices, especially smartphones, has provided significant opportunities to access large-scale personal mobility data (Zheng, 2015). However, the utilization of such rich personal mobility data raises privacy concerns due to the high risk of privacy breaches. Therefore, the generation of pseudo personal mobility data, which can capture real-world mobility patterns without compromising user privacy (Pappalardo & Simini, 2018; Savage, 2023), has emerged as an important task.

Based on in-depth analysis of human mobility features based on millions of real users' mobility data, we find the 'long-tail' issue of human mobility – to restore human mobility, a small number of significant places can reach the passing line, but the number of significant places required for high-quality results increases steeply (*Figure 1*). (Reconstructing 99% static population calls for more than 10 significant places and that number of dynamic populations is 15 or more). And they do not exist in isolation, but rather exhibit coupling relationships within a high-dimensional space. This requires us to generate personalized heterogeneous mobility regularity graph - the 'life patterns' - from the high-dimensional space.

[1]Interfaculty Initiative in Information Studies, The University of Tokyo, Tokyo, Japan
[2]School of Urban Planning & Design, Peking University, Shenzhen, China
[3]Center for Spatial Information Science, The University of Tokyo, Chiba, Japan
[4]Department of Computer and Engineering, Southern University of Science and Technology, Shenzhen, China

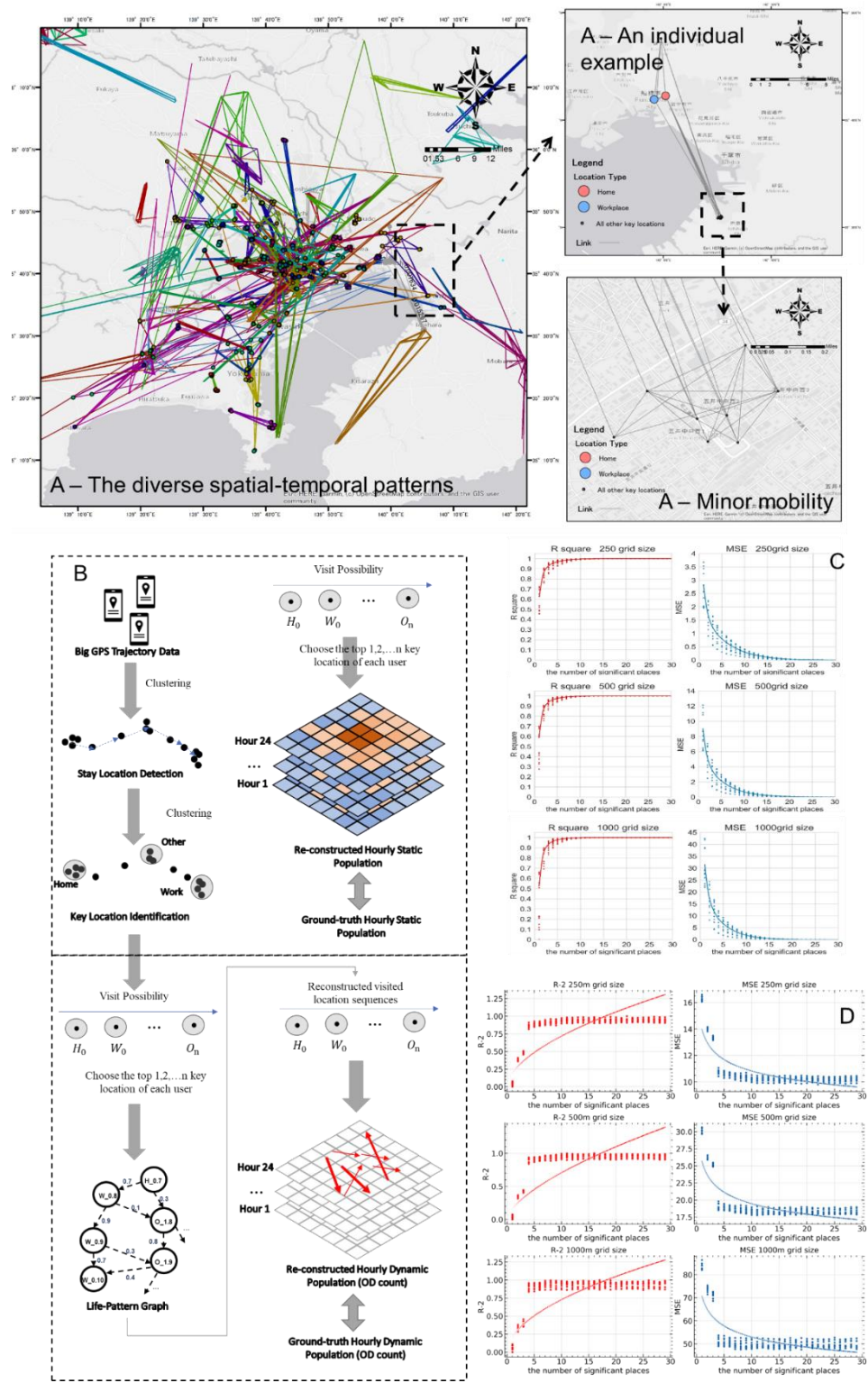

**Figure 1 Highly diverse human mobility and the reconstruction capability of significant places.** A) left part illustrates the diverse – diverse in mobility pattern and spatial space - human mobility of a number of users; A) upper right part shows a typical mobility pattern case of commuter; A) lower

right part shows the partial enlarged detail of the case; B) upper part shows the procedure of hourly static population reconstruction experiment; B) lower part shows the procedure of hourly dynamic OD reconstruction experiment; C) shows the results of static population reconstruction experiment - the average R2 and MSE between the original grid population and the grid population under different sampling conditions. D) shows the results of dynamic population reconstruction experiment - the average R2 and MSE between the original grid population and the grid population under different sampling conditions.

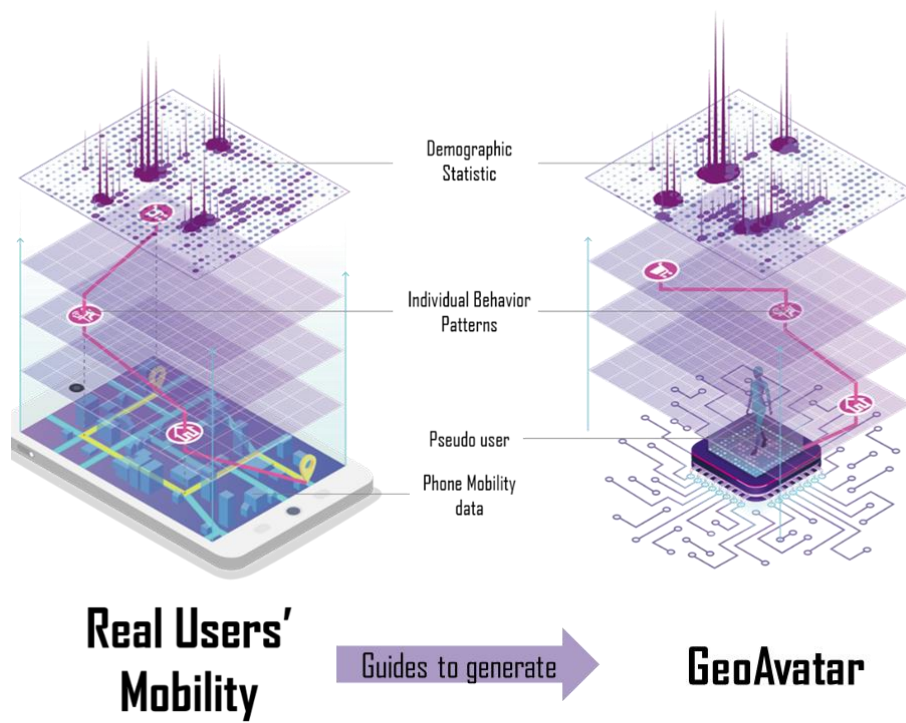

**Figure 2 Big Mobile Phone GPS Data Driven Pseudo Personal Mobility Generator.**

Numerous researchers have investigated pseudo mobility data generation, but gaps still exist. Mainstream approaches for human mobility generation can be classified into two categories: 1) mechanism-based modeling approaches (mainly fully explicit modeling approaches e.g., activity-based methods), such as TimeGeo (S. Jiang et al., 2016), which typically employ regression models to capture human behavior and generate pseudo mobility data based on these models; 2) deep-learning-based methods, including RNN-based models (X. Song et al., 2016; Y. Yu et al., 2019), GAN-based models (W. Jiang et al., 2023; Liu et al., 2018; Ouyang et al., 2018; H. Y. Song et al., 2019), and VAE-based models (Chen et al., 2021; Huang et al., 2019), which directly generate pseudo mobility data in an end-to-end manner. However, both mainstream approaches have their limitations. Activity-based approaches often assume a universal framework for individuals' spatial and temporal decision-making, disregarding the heterogeneity of individual behaviors, resulting in a lack of enough realism. Deep learning-based methods show promise in generating human mobility, but they usually trying trajectory-to-trajectory generation - which also misses the heterogeneity of individuals.

Motivated by our findings (*Figure 1*), this paper introduces an individual-based pseudo human mobility generator called GeoAvatar, which is built upon life patterns (*Figure 2*). GeoAvatar effectively accounts for individual heterogeneity in spatial and temporal decision-making, incorporates demographic characteristics, and provides interpretability.

**Deep Generative Model-driven Life-pattern Generation**

We take the life-pattern generation task as a *graph* generation issue. Graph generation is not a trivial issue due to its complex and variant structure, even with the recent powerful Diffusion-based model (Zhang, M et al., 2023); and different type of graph need targeted consideration. Unfortunately, most of past studies focused on *Molecule Structure* generation (Shi, C et al., 2020), *Protein Design* (Guo, X et al., 2021), and *Social Network* generation (Gao, Y et al., 2018), nearly no studies about life-pattern generation task as a graph generation issue.

Through empirical analysis, we find that: as a kind of *heterogenous graph* (for realistic and diversity it must be heterogenous), life-patterns shows low-dimensional manifolds in a high-dimensional space, whose distribution is hard to captured by ordinary generative model (***Figure 3***). Sampling applying deep generative model like GAN (Generative Adversarial Network) could not be trained for life-pattern generation task. We begin from the original GAN framework, re-extrapolates a series of problems in the original framework, analyzes them in conjunction with the characteristics of the life-pattern generation problem, and finally achieves high-quality life-pattern generation based a Gradient-penalty Wasserstein GAN-like model (*details see Appendix Part*).

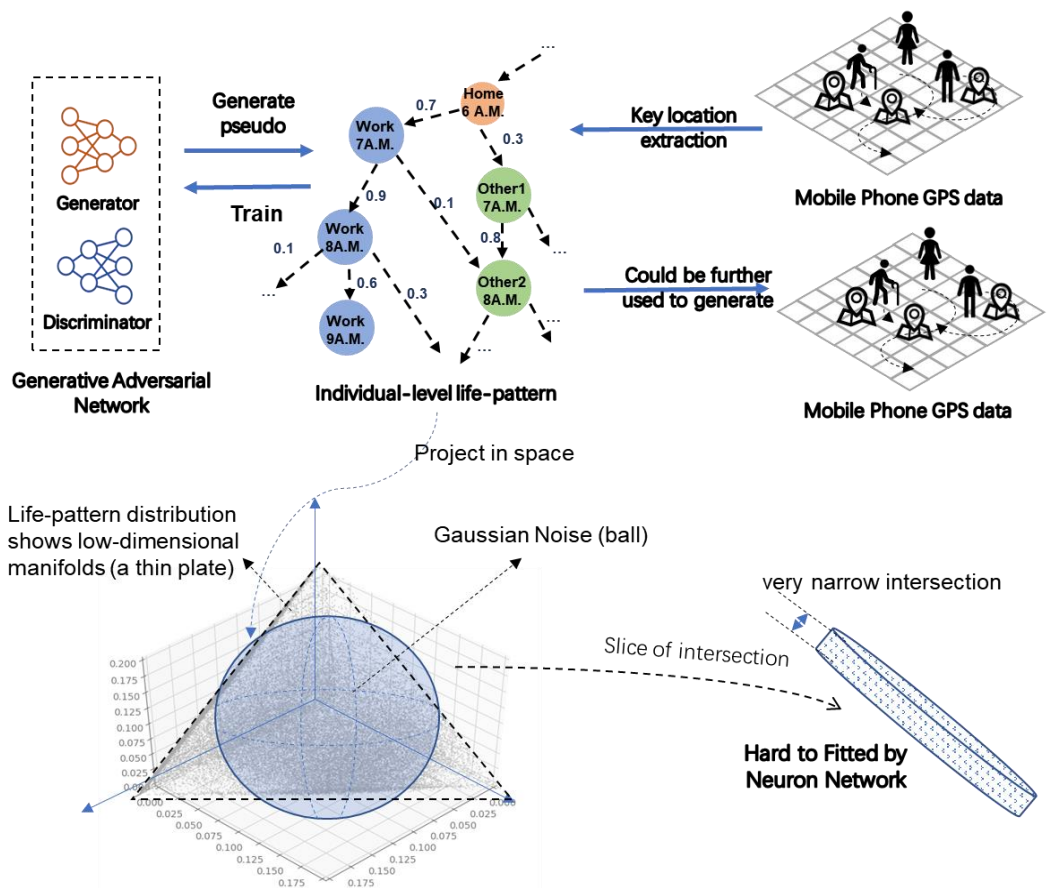

**Figure 3 Illustration of the life-pattern generation** – as a graph generation issue in a high-dimension space. As a kind of heterogenous graph, life-patterns shows low-dimensional manifolds in a high-dimensional space, whose distribution is hard to captured by ordinary deep generative

model.

**Evaluation- Activity Features**

Dressing long-tail issue benefits the reality of activity features (the generated life-patterns), we evaluated the reality of activity features considering three aspects: aggregated average, temporal (hourly) distribution, and individual distribution.

**Aggregated Average in Activity Feature:** To assess the aggregated average, we utilize the Mean Average Error (MAE) of the average activity probability between generated cases and ground-truth cases as an indicator. The GeoAvatar demonstrates better performance compared to the FEM method. The overall MAE of GeoAvatar is 4.54%, whereas the average MAE of FEM is 15.9% (***Figure 4-B***). The superior performance of GeoAvatar can be attributed to its inclusion of the long-tail part. The top frequent key locations, which correspond to specific activities, encompass a large part of daily mobility patterns, but miscellaneous key locations also matter. Neglecting them obviously reduced the average reality - it would not only impact the representation of the miscellaneous long-tail locations themselves but also influence the top frequent key location (e.g., home-stay, work) (as shown in ***Figure 4-A***).

**Hourly Distribution in Activity Feature:** Regarding the hourly distribution, we employ the JS-divergence (Jensen–Shannon divergence) to measure the similarity between pseudos' and trues. GeoAvatar outperforms other methods with a JS-divergence of 0.0356, whereas FEM achieves a value of 0.0972. This indicates that GeoAvatar excels not only in the aggregated average but also in the distribution of hourly probabilities (***Figure 4-C, D*** presents the top 2 frequent cases).

**Individual Distribution in Activity Feature:** We aim for the generated mobility to be both realistic and diverse at the individual level, resembling real-life situations. To evaluate the diversity and similarity of the generated cases, we project the life-pattern graphs from ground-truth and generated data into a 3-dimensional space using NMF (Non-negative Matrix Factorization) (*for more details, refer to Appendix*). We employ JS-divergence as the metric. GeoAvatar exhibits a JS-divergence of 0.047 with the ground-truth life-patterns, whereas the JS-divergence between the FEM method and ground-truth is 0.075. This implies that the life-patterns generated by GeoAvatar have a distribution more similar to the ground-truth, even though GeoAvatar utilizes generated life-patterns while FEM employs real parameters. When observing the scatter plots of the life-patterns, it could be observed that the FEM method lacks diversity in the right part, which represents more-other-place (***Figure 4-E***). The JS-divergence gap between GeoAvatar and FEM reduces compared to the overall results (FEM: 0.044, GeoAvatar: 0.038). This reduction could be attributed to the absence of the long-tail key locations, which are more likely to be categorized as other locations.

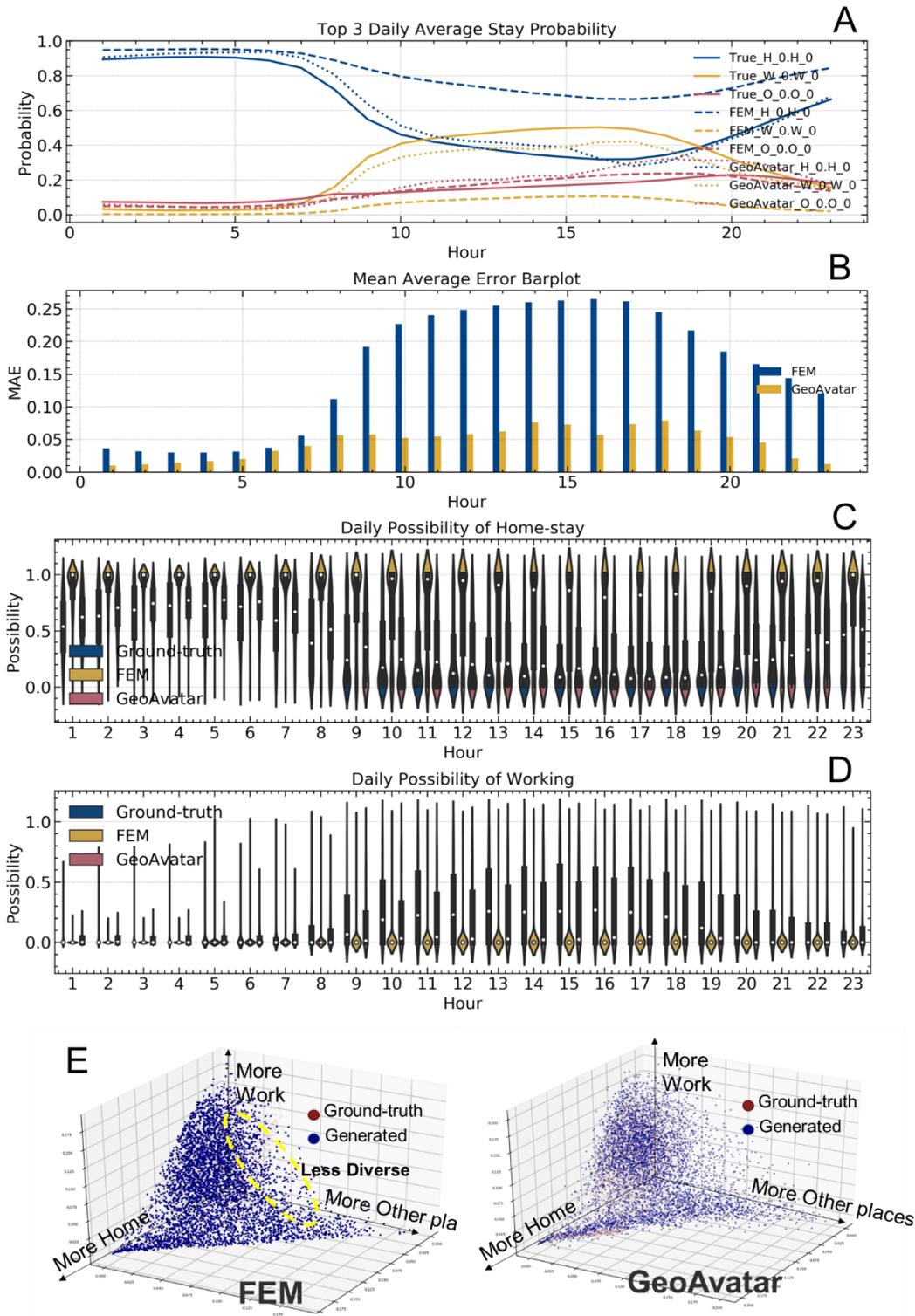

**Figure 4 Activity features comparison.** The x-axis is the 24 hours of a day, the y-axis is the possibility staying in a certain key location. A) shows the daily average probability of frequent activities including home-stay, working and go other places (most-frequent one), the x-axis is the 24 hours of a day, the y-axis is the probability. B) shows the errors of the hourly daily average probability of frequent activities. C) and D) shows the hourly distribution (violin plot) of the hourly daily average probability of frequent activities where C shows the situation of home-stay, and D shows the situation of working. E) shows the life-patterns' distribution comparison in 3-dimension space (each point stands for a life-pattern of a user) where right one is the comparison between GeoAvatar and ground-truth; left one is the comparison between FEM and ground-truth.

Based on the above life-pattern generator, we have developed a comprehensive multi-modal framework called GeoAvatar for generating pseudo personal mobility data. This framework focuses on capturing urban flow spatial patterns and individual behavior patterns, allowing for the simulation of mobile dataset sampling with high spatial and temporal reality.

The GeoAvatar framework consists of four key components, (***Figure2,*** *details refer to Methodology and Appendix*). The PART I component is an individual-level life-pattern generator driven by deep learning model (Generative Adversarial Network). Without a universal This component achieved completely unique, heterogeneous life-pattern graphs for each created pseudo person generation. The PART II component is to perform high-restore mobility sequence re-construction from a given life-pattern. With this component, personal diversity is taken in consideration when re-constructing/generating mobility sequence from life-pattern. The PART III component is a demographic information inference model that could infer the use demographic characteristics based on a given life-pattern, by which each created pseudo person could be assigned a reasonable demographic characteristic. The PART IV part is a key location table generator that could assign suitable geographic coordinates to a given user - considering the inferred demographic information based on a Bayesian network model.

**Evaluation- Physical law**

At a statistical scale, human mobility exhibits patterns that adhere to physical laws, including the distribution of jumping sizes (distance between consecutive locations) and the number of locations visited in a day. These aspects of human mobility follow a universal distribution pattern, and it is important to evaluate how well pseudo mobility data generated conform to these physical laws compared to ground-truth mobility data. To assess this, we employ the Kolmogorov-Smirnov (K-S) statistic as an indicator of the similarity between the distributions.

The results of our evaluation indicate that both the GeoAvatar framework and the baseline FEM (fully explicit modeling, for more details, refer to Baseline Setting and Appendix) method successfully capture the physical laws of human mobility, with no significant performance gap between them. Regarding the jumping size distribution, GeoAvatar achieves a K-S statistic of 0.217, while the FEM method yields a K-S statistic of 0.198 (***Figure 5 A***). In terms of the daily visit location number, GeoAvatar performs slightly weaker but still comparative, with a statistic of 0.440 compared to the FEM method, which obtains a statistic of 0.238 (***Figure 5 B***).

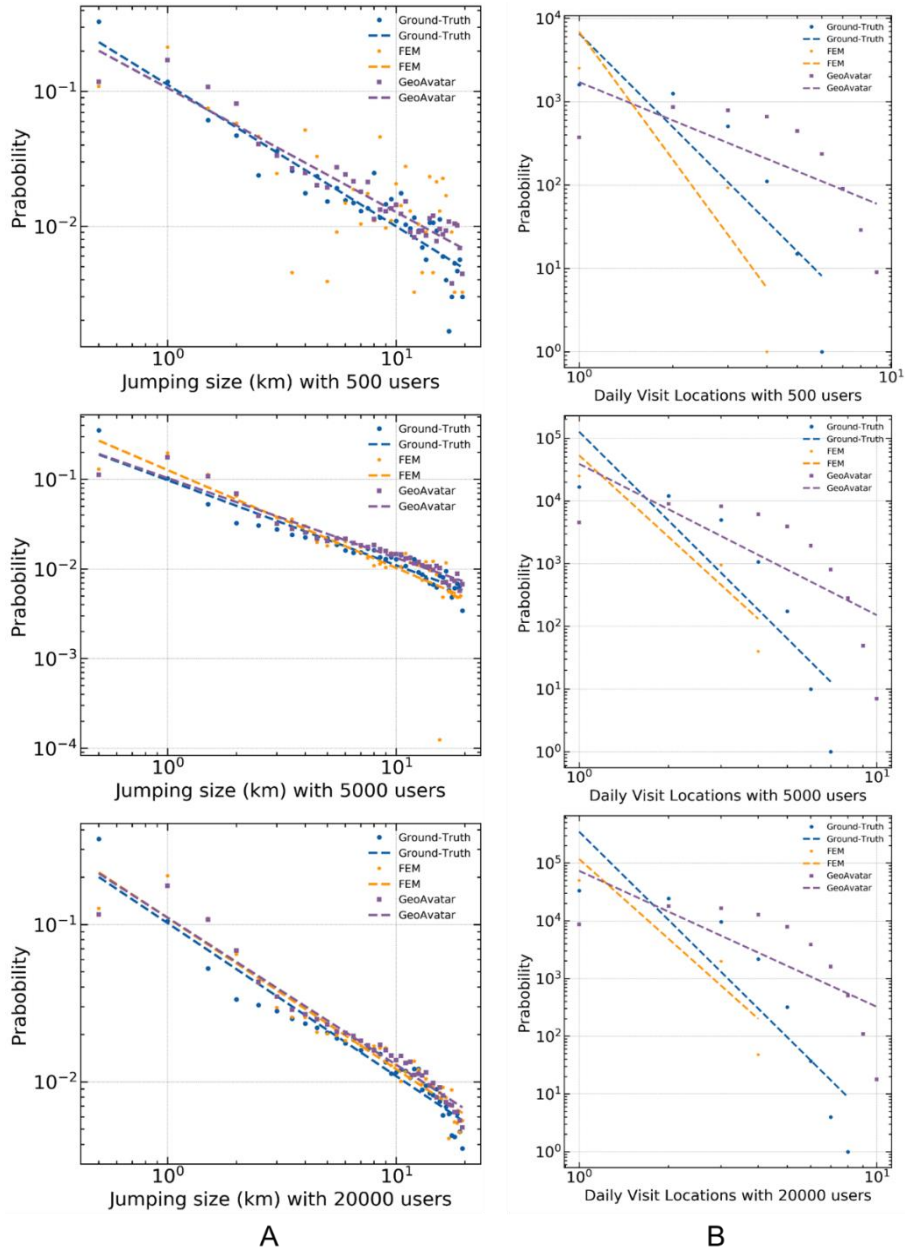

**Figure 5 The performance in term of physical realistic.** A) shows the probability distribution of jumping size under different sampling scale, B) shows the probability distribution of daily visit location number under different sampling scale.

These findings demonstrate that both GeoAvatar and FEM can achieve a comparable level of physical-level realism at the statistical level, which is reasonable to expect. The FEM method utilizes real individual-level temporal and spatial choice parameters, which helps restore the physical laws of human mobility at the statistical level. Moreover, the comparative performance of GeoAvatar showcases its excellent ability to generate pseudo personal mobility data that adheres to physical-level realism, even when starting from pseudo users—a micro-level perspective.

In summary, our evaluation confirms that GeoAvatar successfully captures the physical laws of human mobility and achieves comparable performance to the FEM method. This validates the effectiveness of the GeoAvatar framework in generating realistic pseudo personal mobility data and further supports its utility for various applications in urban studies and beyond.

**Evaluation- Spatial-temporal Features**

For evaluating the spatial-temporal features, we employed two indicators: the correlation (r-square) of grid population and od counts between the generated mobility data and the ground-truth mobility data. Initially, we randomly selected a 7-day trajectory for 60,000 users from the ground-truth data. Due to the variation in human mobility among different users and periods, we divided the ground-truth data for these users into two parts: ground-truth-1 for 30,000 users and ground-truth-2 for the other 30,000 users. Next, we generated a 7-day trajectory for 30,000 users using the proposed method and baseline methods. We then aggregated the trajectories into grids and computed the hourly grid population and hourly od population for both the generated and ground-truth data. Finally, we calculated the r-square as the evaluation indicators (*details refer to Appendix*).

**Tokyo Case**. In Tokyo case, the r-square value for the grid population between ground-truth-1 and ground-truth-2 data is 0.895. The r-square value for the grid population between ground-truth-1 and the data generated by GeoAvatar is 0.896, while for FEM it is 0.492, and for TimeGeo it is 0.370. Regarding the od count, the r-square value between ground-truth-1 and FEM is 0.611, while for TimeGeo it is 0.671, and for GeoAvatar it is 0.832 (slightly lower than the natural fluctuation) (***Figure 6***, for figure of TimeGeo, refer to *Appendix*). Similarly, we conducted experiments in Beijing and Shanghai, we also found similar performance.

**Beijing Case.** For the Beijing case, analyzing the performance of GeoAvatar, the r-square value for the grid population between ground-truth-1 and the data generated by GeoAvatar is 0.821 for the Beijing GeoAvatar error and 0.638 for the od r-square. In comparison, TimeGeo shows a r-square value of 0.576 for the Beijing TimeGeo grid population error and 0.440 for the od r-square. Concerning natural fluctuation, the comparison between ground-truth-1 and ground-truth-2 data yields an r-square value of 0.867 for grid population and 0.716 for od count.

**Shanghai Case.** In the Shanghai case, GeoAvatar's performance is indicated by an r-square value of 0.729 for the Shanghai GeoAvatar grid population error and 0.687 for the od r-square. TimeGeo's performance is represented by a r-square value of 0.496 for the Shanghai TimeGeo grid population error and 0.465 for the od r-square. For natural fluctuation, the r-square values are 0.840 for grid population and 0.713 for od count, based on the comparison between ground-truth-1 and ground-truth-2 data.

Based on these results, it can be concluded that GeoAvatar achieves high accuracy in spatial-temporal demographics, while FEM and TimeGeo methods fail to do so. The results highlight the importance of utilizing a heterogeneous life-pattern structure and key locations, which allow for the exploration of more complex structures while maintaining spatial realism. While human mobility can be well-reconstructed with a few key locations (***Figure 1***), it is crucial to consider the long tail part. Neglecting this aspect can influence all steps, from mobility pattern construction and mobility sequence sampling to key location sampling. The accumulation of errors ultimately leads to significant failures in achieving demographic statistical accuracy (***Figure 6***). Additionally, mechanism-based models like TimeGeo typically employ an EPR-like approach that samples locations based on rules, leading to biases in temporal-spatial population representation - these models fail to consider that location choice is a series of combinations for a specific user, rather than just a mechanism-based random traveling.

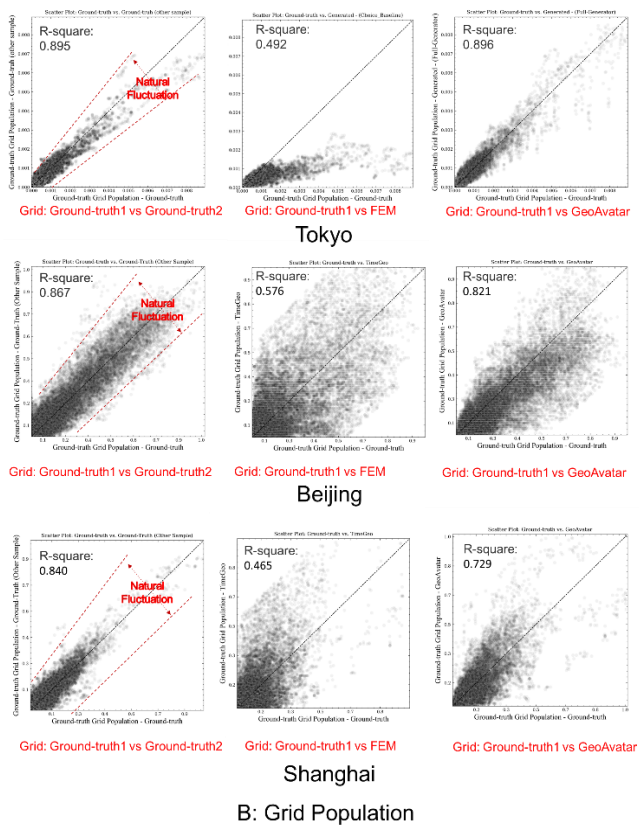
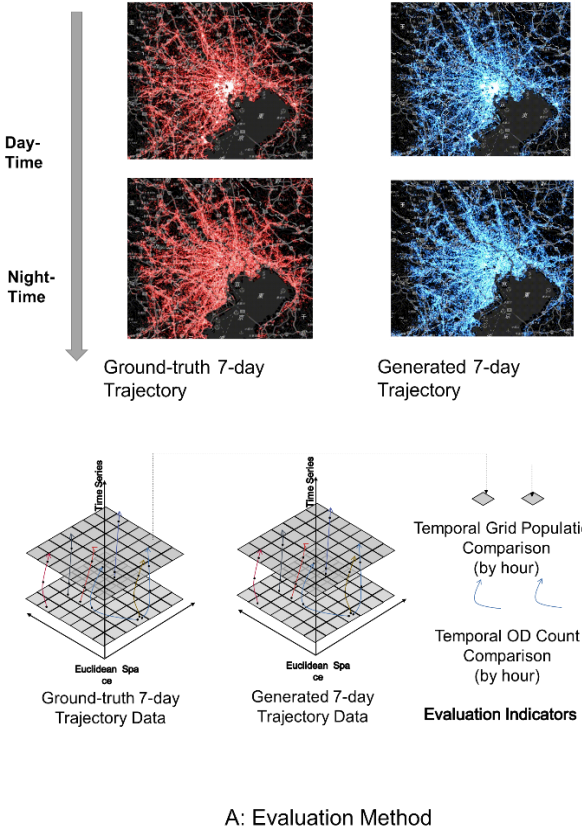
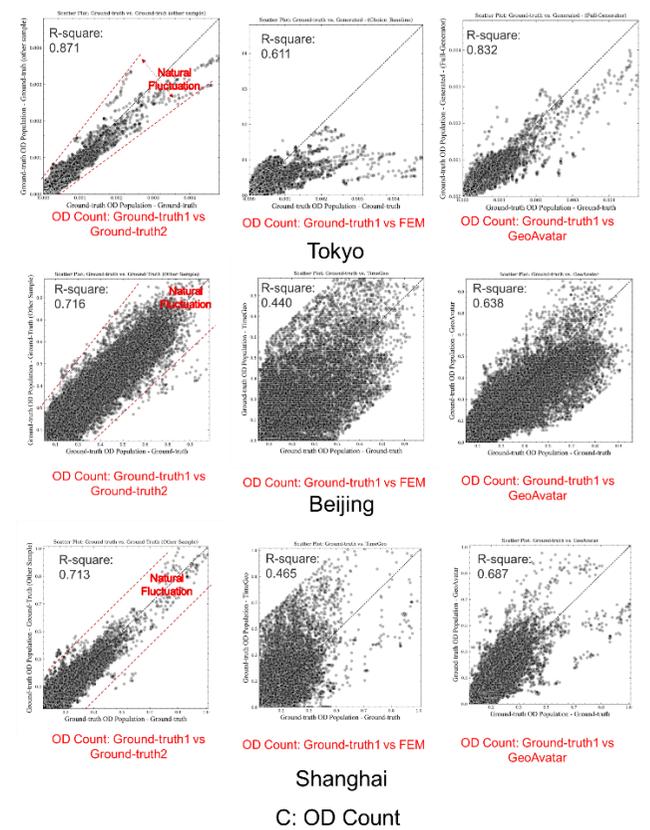

B: Grid Population     A: Evaluation Method     C: OD Count

**Figure 6 Evaluation of spatial-temporal Features.** A) shows the evaluation method and metrics. B) The overall comparison of grid-population scatter plots between ground-truth-1 and ground-truth2 the ground-truth-1 and FEM's generation, and the ground-truth-1 and GeoAvatar's generation in Tokyo, Beijing and Shanghai. C) The overall comparison of od-count scatter plot between ground-truth-1 and ground-truth2 the ground-truth-1 and FEM's generation, and the ground-truth-1 and GeoAvatar's generation in Tokyo, Beijing and Shanghai.

**Conclusion**

We developed GeoAvatar, an individual-based pseudo human mobility generator designed to produce highly realistic pseudo personal mobility data. The proposed method encompasses a deep generative model that generates heterogeneous individual life-pattern graphs from Gaussian noise, a reliable demographic labeler that infers pseudo users' demographic characteristics, and a Bayesian/Markov-based key location table generator that determines key location choices based on demographic information.

The evaluation of the proposed method focuses on three aspects: physical law, activity features, and spatial-temporal features. Firstly, the physical law evaluation compares the jumping size and stay duration distributions of the generated data with those of the ground truth data using the KS-test. The results demonstrate that the performance of the proposed method is comparable to that of the baseline methods. Secondly, the activity feature evaluation involves analyzing the aggregated average probabilities for activities, the hourly distribution of activity probabilities, and the distribution of individual life-patterns. GeoAvatar exhibits higher accuracy in overall life-patterns and better diversity compared to the baseline method. Lastly, the spatial-temporal feature evaluation compares the grid population and od count correlation (r-square) of the generated data with those of the ground truth data. The results reveal that GeoAvatar significantly outperforms the baseline methods in terms of macroscope realism, with notably higher r-square values in population and od count comparisons.

In summary, GeoAvatar successfully achieves powerful capabilities to generate realistic pseudo personal human mobility data, with no accessing to true individual-level personal information. Furthermore, the proposed method maintains extensibility for broader applications, holding significant promise as a new paradigm for generating human mobility data.

**Methodology**

The GeoAvatar consists of four key components:

   **Life-Pattern Generator**: Recognizing the long-tail nature of human mobility reconstruction, we acknowledge that a rule-based method with a preset unified framework alone cannot achieve highly accurate pseudo personal mobility generation. To address this, we employ a deep learning approach utilizing generative adversarial networks (GANs). This deep generative model enables the generation of individual-scale and heterogeneous life-patterns, capturing the diverse mobility patterns of individuals (for more details, refer to Appendix Chapter 3).

   **Human Mobility Sequence Generator**: While life-patterns provide a foundation for reconstructing individual activity sequences, relying solely on random walks on the life-pattern graph is insufficient. We propose the Graph-walk With a Guide (GWG) method. This approach utilizes a deep-learning-based model to guide the individual activity sequences reconstruction from life-pattern (i.e., fine tuning the probabilities of sampled mobility sequences), to generate more realistic human mobility sequences. By incorporating preceding information and sampling probability correctors, the GWG method enhances the accuracy of pseudo personal mobility generation (for more details, refer to Appendix Chapter 4).

   **Demographic Information Labeler**: The third component of our pseudo personal mobility generator is a model for inferring demographic information based on an individual's life-pattern. While it has been observed that life-patterns are associated with demographic characteristics, using individual-level ground-truth demographic labels raises privacy concerns. To overcome this challenge, we propose a trustworthy demographic labeling model that leverages mobility data and census data. Our approach utilizes a meta-graph-based data structure to represent users' life-patterns and projects them into a low-dimensional feature space. Based on these features, we employ a variation-inference-based advanced Bayesian model for demographic inference. Then, when we generated a pseudo life-pattern, we could label the demographic information for the pseudo user (for more details, refer to Appendix Chapter 5).

   **Key Location Table Generator**: Considering the limitations of EPR (exploration potential ratio) methods, which tend to rely heavily on random exploration and overlook individual attributes, we introduce a Markov method based on pseudo person attributes. This method, labeled by the Demographic Information Labeler, generates a key location table for pseudo persons. The key location table serves as a geographic location correspondence table, incorporating individual attributes and enhancing the realism of the generated pseudo personal mobility data (for more details, refer to Appendix Chapter 6).

   In summary, a pseudo individual life-pattern graph is generated from a deep generative network. Based on the individual life-pattern graph, demographic information is inferred, and human mobility sequence is sampled from the pseudo life-pattern. Then, based on a Bayesian network model, combining the inferred demographic information, we generated key location table for each pseudo individual. Finally, the pseudo personal mobility data is generated.

**Baseline Settings**

To confirm the ideas driving the GeoAvatar, we employ two approaches as the baselines: FEM(Full-Explicit Modeling) – what if we use real parameters of activity model but limited key location number, and TimeGeo – does EPR-like model could re-construct spatial-temporal people flow well?

**FEM(Full-Explicit Modeling)**: Start with a large dataset of mobile phone GPS data, we utilize clustering algorithm to identify stay points, which are locations where users spend a significant amount of time. Then, we identified significant places (e.g., home, workplaces, other frequently visit daytime spots, other frequently visit nighttime spots and others.) by clustering the stay points for each user. And a series of significant places of a user consist of the key location of him/her. These key locations serve as anchor points for generating synthetic mobility patterns. Second, we construct a Bayesian model for each user, capturing their decision-making process based on historical mobility patterns, with limited key location type (home, workplace, top frequent other only). Given any behavior in the previous time point of a user, the probability of behaviors at each time point is stored. Lastly, we generate synthetic mobility patterns for each user by simulating sequences of actions and locations. Utilize the individual-specific Bayesian probabilities to determine the choices made by users at each time point, and taking the true key location of this user, the mobility data is generated/re-constructed (for more details, refer to Appendix Chapter 7).

**TimeGeo**: Based on the classical TimeGeo framework, we fitting the parameters by our dataset, then generate pseudo human mobility data (for more details, refer to *Appendix Chapter 8*).